\documentclass[twocolumn,showpacs,preprintnumbers,amsmath,amssymb]{revtex4}

\usepackage{graphicx}
\usepackage{dcolumn}
\usepackage{bm}
\usepackage{textcomp}

\begin{document}


\title{Fracture of a viscous liquid}

\author{\'Elise Lorenceau}
\affiliation{Physique de la Mati\`ere Condens\'ee, UMR 7125 du
CNRS, Coll\`ege de France,\\ 75231 Paris Cedex 05, France. }
\author{Fr\'ed\'eric Restagno}
\author{David Qu\'er\'e}
\affiliation{Physique de la Mati\`ere Condens\'ee, UMR 7125 du
CNRS, Coll\`ege de France,\\ 75231 Paris Cedex 05, France. }

\date{\today}

\begin{abstract}
When a viscous liquid hits a pool of liquid of same nature, the
impact region is hollowed by the shock. Its bottom becomes
extremely sharp if increasing the impact velocity,  and we report
that the curvature at that place increases exponentially with the
flow velocity, in agreement with a theory by Jeong and Moffatt.
Such a law defines a characteristic velocity for the collapse of
the tip, which explains both the cusp-like shape of this region,
and the instability of the cusp if increasing (slightly) the
impact velocity. Then, a film of the upper phase is entrained
inside the pool. We characterize the critical velocity of
entrainment of this phase and compare our results with recent
predictions by Eggers.
\end{abstract}

\pacs{47.15.Rq, 47.15.Gf, 47.55.Dz}%
\maketitle

We all observed that diving in a swimming-pool produces air
bubbles, while entering slowly in the same pool does not: the
motion of a solid penetrating a liquid at a high speed largely
deforms the surface of this liquid, and induces air entrainment.
This phenomenon often defines the maximum speed of coating of a
solid, which enters a bath before leaving it coated. Generally air
entrainment must be avoided because of the resulting bubbles
(which are all the more harmful since the coating solution
generally contains surfactants, and thus is likely to transform
into a foam). Different experiments using either fibers or plates
as solids indeed prove the existence of a threshold (in velocity)
for air entrainment \cite{Burley92}, but there is today no
theoretical picture for understanding quantitatively this
phenomenon.

Quite similarly, when a viscous liquid enters a pool of same
nature (as it occurs for example for the stem wave of a moving
ship), air entrainment may occur (producing bubbles, which
ultimately burst -- a serious cause of noise for a ship). Such a
bubble production was observed for a liquid jet impinging a bath
of same nature, as reported by Lin and Donelly, who studied
experimentally the minimum velocity of air entrainment as a
function of different characteristics of the jet (diameter,
viscosity or surface tension) \cite{Lin81}. This question is
important in many industrial processes where a viscous liquid
(typically molten glass, metal or polymer) is poured inside a
mould -- and there again, the formation of bubbles is detrimental.
A remarkable geometry for studying this phenomenon, and more
generally the phenomenology of the regimes of deformation of the
liquid bath, was proposed by Joseph \cite{Joseph91,Joseph92}. Two
counter-rotative cylinders are partially immersed in a bath, whose
upper surface is observed. At a low speed of rotation, the
interface between the cylinders is hardly deformed, but increasing
this speed leads to the formation of a cusp-like shape. This
device thus produces a singularity at a free interface, and more
generally the observation and description of similar singularities
has been a very active field for the past years, either
theoretically \cite{Jeong92} or experimentally - let us quote for
example selective withdrawal \cite{Cohen2002}, the pinch-off of a
jet \cite{Cohen99}, drops running-down on a tilted solid
\cite{Podgarski2001} or jet eruptions \cite{Zeff2000}.

Surface tension naturally opposes the formation of such
singularities at a liquid/fluid interface, and for the problem of
the cusp between two counter-rotative cylinders, Jeong and Moffatt
calculated in a classical paper the (non-zero) radius $r$ of the
tip, in the limit of small Reynolds numbers \cite{Jeong92}. They
derived an analytical expression, which can be summarized using a
physical argument proposed by Hinch \cite{Jeong92}. In this
two-dimensional geometry, the tip can be seen as a thin
(hemi-)cylinder, of radius $r$. The viscous drag (per unit length)
on a cylinder scales as $\eta V/\ln(d/r)$, denoting $d$ as an
external length, $\eta$ as the liquid viscosity and $V$ as the
flow velocity around the cylinder. On the other hand, the surface
tension $\gamma$ draws this cylinder (as it does on the edge of a
free sheet of liquid) with a force $2\gamma$ (there again, per
unit length). Balancing both these forces yields as a tip radius:
\begin{equation}\label{equ_moffat}
    r  \sim  d \exp(-Ca)
\end{equation}
where $Ca$ is the capillary number ($Ca = \eta V/\gamma$).

This law first helps to understand why a tip forms around a
well-defined velocity: the cusp becomes very sharp when $V$ is of
the order of $\gamma/\eta$, a logical quantity since both these
parameters play antagonist roles in the formation of the tip. But
it also predicts that $r$ essentially vanishes above this
velocity, and Eggers pointed out that this should lead to the
crack of the tip, because of the lubrication pressure of the upper
fluid, forced to flow in this very confined region
\cite{Eggers2001}. He showed that beyond a critical radius for the
tip, the pressure in the cusp breaks its stationary shape. Thus,
this critical radius $r_c$ must increase with the viscosity
$\eta_0$ of the upper fluid, and Eggers found that $r_c$ should
scale as $(\eta_0/\eta)^{4/3}$ \cite{Eggers2001}. Together with
Moffatt's result (Eq. \ref{equ_moffat}), this yields as a
threshold velocity $V_c$ of entrainment of the upper phase:
\begin{equation}\label{eq_eggers}
    V_c  \sim \frac{\gamma}{\eta}\ln\left(\eta/\eta_0\right)
\end{equation}
 We could have expected (naively) that entrainment of the upper
fluid would occur when the viscous forces overcome capillary ones
({\em i.e.} $V_c \sim \gamma/\eta$). Eggers' theory (Eq.
\ref{eq_eggers}) shows that taking into account the flow of the
upper phase in the cusp should induce a logarithmic correction to
this simple statement. As a consequence, the threshold velocity
should depend on the viscosity of the upper phase, {\em i.e.} on
its nature. This might sound {\em a posteriori} quite logical, but
remarkably, this dependence is expected to be extremely weak.

Here, we investigate experimentally the different situations
related to this problem. We first describe the macroscopic
deformation of the interface, due to the incoming flow of liquid.
Then, we focus on the smaller scale of the tip. Finally, we
characterize the condition of entrainment of the upper phase.

\begin{figure}[htbp]
\includegraphics[height=3cm]{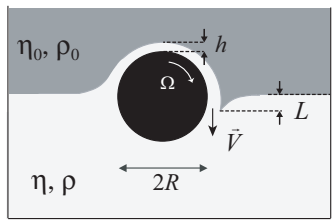}
\caption{Experimental set-up. A solid cylinder is placed at the
interface between a viscous dense liquid and a lighter fluid. The
deformation of the interface between both fluids is monitored as a
function of the speed of rotation of the
cylinder.}\label{fig_setup}
\end{figure}

We used a horizontal cylinder of radius $R = 2$~cm, half-immersed
in a bath of glycerol (or in some cases silicone oil), of density
$\rho$ and viscosity $\eta$ around 1 Pa.s (Figure
\ref{fig_setup}). The whole device was set in a transparent tank,
and the upper fluid was either air or light oil
(of density $\rho_0 < \rho$ and viscosity $\eta_0\ll\eta$). The
interfacial tension between the two fluids, denoted as $\gamma$,
was measured for each experiment. The cylinder could be rotated at
a controlled velocity $\Omega$, and the motion filmed with a
digital camera.

\begin{figure}[htbp]
\includegraphics[width=8.7cm]{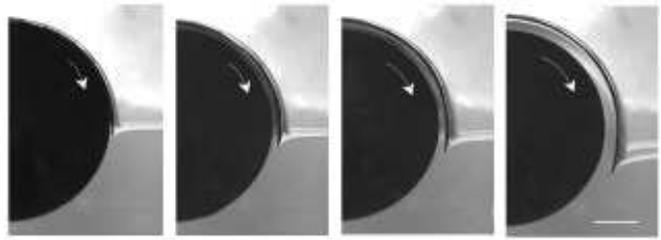}
\caption{Four snapshots showing the bath deformation for an
increasing speed of rotation of the cylinder ($V$ is 6~cm/s,
10~cm/s, 17~cm/s and 30~cm/s, going from left to right). The lower
liquid is a silicon oil of viscosity 0.97~Pa.s and the upper fluid
is air. The bar in the last picture indicates
1~cm.}\label{fig_4snapshots}
\end{figure}

A viscous liquid film of thickness $h$ (of typically 5~mm, thus
thick enough to be independant of the nature of the roller
surface) is dragged out of the bath. This layer impinges the bath
at a velocity $V \simeq \Omega(R+h)$, and deforms the free surface
at this place by a depth $L$. We denote $r$ as the radius of
curvature of the bottom of this hollow region. Figure
\ref{fig_4snapshots} shows the different shapes obtained with a
silicon oil ($\eta = 0.97$~Pa.s), and increasing the speed of the
cylinder.

It is observed that both the characteristic lengths $L$ and $r$
are dynamic in essence: the depth of immersion $L$ increases with
the speed of cylinder, while the tip radius $r$ decreases.
Moreover, these lengths are very different from each other: while
the depth of immersion is around 1~cm, the tip has a radius of
curvature much smaller than 1~mm. This radius can become very
small when the speed is large, as first reported by Joseph et al.
\cite{Joseph91} in a similar situation. But we do not observe as
they did a clear threshold velocity for the formation of a cusp --
both the lengths $L$ and $r$ varying continuously as a function of
the cylinder velocity.

\begin{figure}[htbp]
\includegraphics[width=8cm]{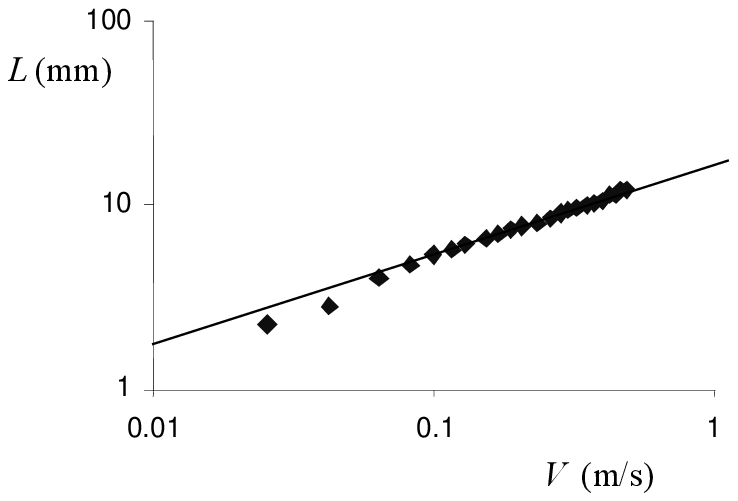}
\caption{Depth of immersion of the tip as a function of the
velocity of the incoming liquid. The bath is a viscous silicon oil
($\eta = 0.97$~Pa$\cdot$s) with air above. The straight line
indicates a slope 0.5.}\label{fig_ldependsV}
\end{figure}

We first focus on the macroscopic length $L$, namely the depth of
immersion of the tip into the bath, and report in Fig.
\ref{fig_ldependsV} the variation of $L$ as a function of the
impacting velocity $V$, for a bath of silicon oil ($\eta =
0.97$~Pa.s) .

 The depth $L$ of immersion of the tip is found to increase with $V$. For
speeds larger than 10~cm/s, $L$ is several millimetres, which is
significantly larger than the capillary length
$\kappa^{-1}=\sqrt{\gamma/\Delta\rho g}\approx 1.5$~mm (denoting
$\Delta\rho$ as the difference of density between both fluids). We
conclude that capillary forces can be neglected in this regime of
large velocities. The viscous friction which pulls downward the
tip can be written dimensionally $\eta V/L^2$, while gravitational
forces per unit volume (which pull it upwards) are $ \Delta\rho
g$. Balancing both these forces yields:
\begin{equation}\label{equ_longvisqueuse}
L\sim\left({\eta V}/{\Delta\rho g}\right)^{1/2}
\end{equation}

Equation \ref{equ_longvisqueuse} defines some kind of {\em viscous
length} (on the model of $\kappa^{-1}$ defined just above,
replacing capillary effects by viscous ones). It describes fairly
well the data in Figure \ref{fig_ldependsV}, where $L$ is observed
to increase as the square root of $V$. Some deviations appear at
small $V$ : the deformation then is small ($L \sim \kappa^{-1}$),
and capillary effects must be considered. Since they oppose the
formation of the hollow region, they indeed make $L$ smaller than
predicted by Eq. \ref{equ_longvisqueuse}.

\begin{figure}[htbp]
\includegraphics[width=8.5cm]{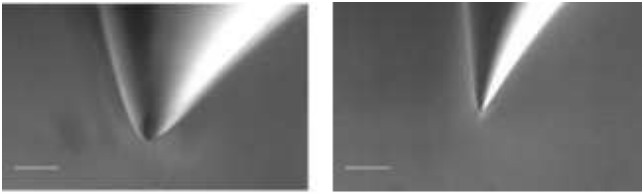}
\caption{Macrophotograph of the tip region (the bar indicates 200
\textmu m), for a tip of air in a bath of glycerol. The flow
velocities are respectively 14 cm/s and 22 cm/s, and the sharpness
of the tip is observed to be highly dependant on this velocity.
}\label{fig_tip}
\end{figure}

We then focused on the small scale of the deformation, namely the
tip itself, and took macrophotographs of this region. A numerical
camera was placed behind an optical microscope lighted by a
classical white lamp, and the whole device first calibrated using
micrometric  fibers. The tip was observed to be slightly deeper
close to the container walls (which allowed us to make
conveniently our measurements), but remains far from it (about 15
mm because of the lubrification of the wall. As reported in Figure
\ref{fig_tip}, the tip has a typical radius smaller than
100~\textmu m, which decreases strongly with the flow velocity.

\begin{figure}[htbp]
\includegraphics[width=6cm]{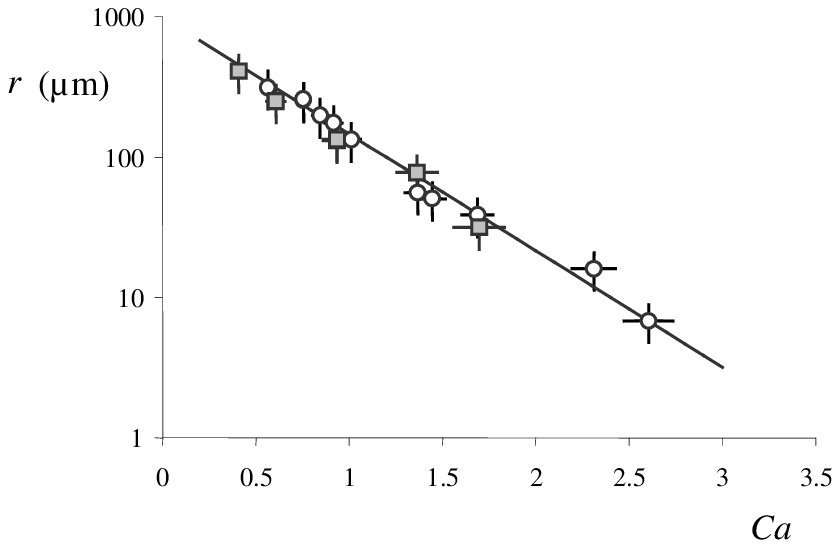}
  \caption{Tip radius as a function of the capillary number
  $Ca = \eta V/\gamma$ for air tips formed in a bath of glycerol (circles, $\eta= 0.91$~Pa$\cdot$s, $\gamma = 55$~mN/m),
  and oil tips in a bath of glycerol (squares, $\eta = 0.53$~Pa$\cdot$s, $\eta_0 = 0.5$~mPa$\cdot$s, $\gamma = 22$~mN/m).
  The data are deduced from macrophotographs such as the ones in Figure \ref{fig_tip}. The plot is semi-logarithmic,
  and the equation of the straight fitting line is $r = 993 \exp(-1.91 Ca)$ (expressed in micrometers).}\label{fig_moffat}
\end{figure}

Analyzing the contour of the interface close to the tip allows a
determination of the curvature: we can fit this contour by a
parabolic function $y = ax^2$, from which we deduce the curvature
$r = 1/2a$. We display in Figure \ref{fig_moffat} the measured tip
radii as a function of the capillary number $Ca = \eta V/\gamma$
for air and oil tips formed in a bath of glycerol.

\begin{figure}[htbp]
\includegraphics [height=3.3cm]{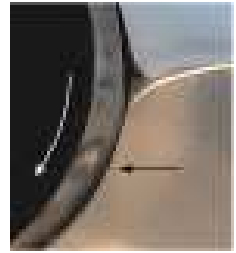}
\caption{Above a well-defined velocity $V_c$, a thin yet visible
sheet of the upper fluid (here air), appearing in black in the
image and stressed by an arrow, comes out of the tip and enters
the viscous bath (here glycerol). The velocity $V$ is 32~cm/s,
while the threshold $V_c$ for air entrainment is 30
cm/s.}\label{fig_airentrainment}
\end{figure}

The tip radius is found to decrease in a continuous way, as a
function of the flow velocity $V$. More precisely, its logarithm
varies linearly with $V$ -- in agreement with Moffatt's law (Eq.
\ref{equ_moffat}), on nearly two orders of magnitude for the tip
radius. This exponential behavior helps understanding why the
cusp-like shape apparently sets for a well-defined velocity, of
the order of $\gamma/\eta$, although the actual variation of the
radius is continuous with $V$. But it also suggests that slightly
above this velocity, the cusp should become extremely sharp. This
is not the case: a small increase of the roller velocity leads to
the destruction of the tip, and the upper phase then is entrained,
as reported in Figure \ref{fig_airentrainment}. Note that the
entrainment occurs simultaneously all along the tip, showing the
2D nature of the phenomenon.

\begin{figure}[htbp]
\includegraphics[width=8cm]{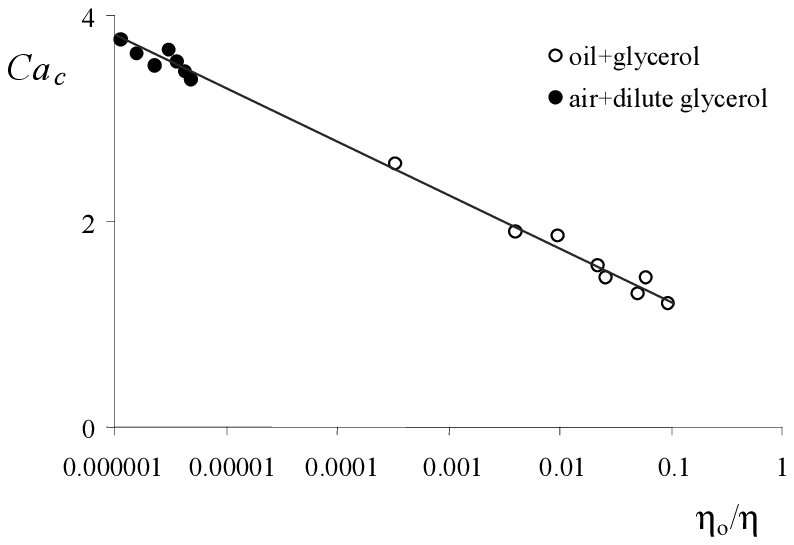}
  \caption{Critical capillary number $Ca_c=\eta V_c/\gamma$ above which the
upper phase is entrained in the viscous bath as a function of the
ratio between both viscosities (semi-logarithmic scale). The full
points are obtained using water-glycerol mixtures (250 mPa$\cdot$s
$<\eta<$1500 mPa$\cdot$s)with air above, while the open points
correspond to glycerol ($\eta = 900$~mPa$\cdot$s) with above
different light oils (0.3~mPa$\cdot$s $<\eta_0<
$83~mPa$\cdot$s).}\label{fig_eggers}
\end{figure}

We denote $V_c$ as the threshold velocity and $Ca_c$ as the
corresponding capillary number. Our experiments consisted in
determining $Ca_c$. Firstly, we varied the bath viscosity $\eta$,
using mixtures of water and glycerol and air as the upper fluid.
$\eta$ was measured for each solution. The corresponding data are
displayed in black in Figure \ref{fig_eggers}. Secondly, we used
pure glycerol for the bath, and various light oils (such as
hexane, hexadecane and silicone oils) as the upper phase. We
measured $\gamma$ for each oil (it was found to be quite constant,
around $26 \pm 1$~mN/m). The corresponding critical capillary
numbers are represented by white points, in Figure
\ref{fig_eggers}, where all our results are plotted as a function
of the viscosity ratio $\eta_0/\eta$, in a semi-logarithmic scale.

We clearly observe an effect of the viscosity $\eta_0$ of the
upper fluid (open points): the larger $\eta_0$, the smaller the
critical capillary number. Nevertheless, this dependency is weak,
since it appears to be well described by a logarithmic law. It is
also limited: if the upper fluid is too viscous (typically
$\eta_0/\eta>0.1$), the lower liquid is not entrained on the
roller. For a given upper fluid, varying the bath viscosity $\eta$
also leads to a (weak) variation of the threshold in capillary
number -- which thus is not simply given by a value of order
unity. All the data lie on the same semi-logarithmic plot, in
agreement with Eggers' model (Eq. \ref{eq_eggers}), for a
variation on five orders of magnitude of the viscosity ratio $
\eta_0/\eta$. However, the slope deduced from the fit is $-0.22$,
larger than $-1$ the value predicted by Eq. \ref{eq_eggers}.

On the whole, these experiments may help clarifying the question
of the stationary pattern which sets as a viscous liquid hits
itself, in a two-dimensional geometry. If the impinging velocity
is large enough ({\em i.e.} for capillary numbers typically larger
than 0.1), two very different dynamic lengths can be defined: a
global one, which is the depth of liquid hollowed by the impact,
and which results from a balance between viscous effects and
gravity; and a local one, at the bottom of this hollow region
which can be expressed as the radius of curvature at this place.
As predicted by Moffatt \cite{Jeong92}, this radius decreases
exponentially with the capillary number. This law defines a
characteristic velocity for which the hollow region becomes very
sharp, {\em i.e.} takes a cusp-like shape, as first reported by
Joseph \cite{Joseph91}. At the same time, it predicts a collapse
of the tip radius for finite velocities, which implies the crack
of the tip (entrainment of the upper phase), as postulated by
Moffatt \cite{Jeong92} and shown theoretically by Eggers
\cite{Eggers2001}. We indeed report here such a crack, and show
that the critical velocity for which it occurs is mainly governed
by the bath viscosity, but also depends (in a much weaker manner)
on the viscosity of the upper phase -- in agreement with Eggers'
theory.

Different studies could naturally complement this one. Firstly, it
would be worth understanding (both experimentally and
theoretically) the thickness of the film of air, in the regime of
entrainment. The selection of this thickness could remind the
Landau-Levich problem (selection of the thickness of the film
adhering to a solid withdrawn out of a bath), but the problem here
is purely dynamical, which raises an interesting (and open)
question (as in selective withdrawal, for the radius of the
emitted filament). The problem of the stability of this film,
which eventually breaks in bubbles, also remains to be studied.
Another issue concerns the practically important question of
water, for which the threshold velocity is much higher, which
would require to study the influence of inertia in all the
different processes described all along this study. Secondly, we
observed a deviation towards the law displayed in Figure
\ref{fig_eggers} using a very viscous silicone oil as a bath
(instead of glycerol), and air above. Then, for $\eta > 10^5
\eta_0$, we measured a critical capillary number much higher,
typically of the order of 5 to 10 (and increasing rapidly as
decreasing the ratio $\eta_0/\eta$). A possible cause of this
discrepancy could be the polymeric nature of the liquid (for which
the chains are all the longer since the viscosity is high), which
could lead to non-Newtonian effects close to the tip - and thus to
changes in the threshold of air entrainment. Using newtonian
silicone oils of smaller viscosity (with either air or ethanol
above) indeed yields data in very close agreement with the ones
displayed in Figure \ref{fig_eggers}. Finally, it would be useful
to connect this study with similar ones in other dimensionality.
In particular, Cohen stressed that for selective withdrawal ({\em
i.e.} one-dimensional tip forming when sucking locally a liquid
above an interface), the viscosity of the fluid forming the tip
does not affect the value of the cut-off curvature
\cite{Cohen2002b}. This could be due to the fact that the Stokes
law on a (hemi-)sphere does not contain any logarithmic
dependency, as it does for a (hemi-)cylinder, which finally
emphasizes that the effects reported here are two-dimensional in
essence.

 {We thank Jens Eggers for stimulating discussions,
Ghislaine Ducouret who helped us in the viscosity measurements and
Gilles Jandeau and Daniel D\'etry for their technical help.}

\printfigures

\begin{thebibliography}{11}
\expandafter\ifx\csname
natexlab\endcsname\relax\def\natexlab#1{#1}\fi
\expandafter\ifx\csname bibnamefont\endcsname\relax
  \def\bibnamefont#1{#1}\fi
\expandafter\ifx\csname bibfnamefont\endcsname\relax
  \def\bibfnamefont#1{#1}\fi
\expandafter\ifx\csname citenamefont\endcsname\relax
  \def\citenamefont#1{#1}\fi
\expandafter\ifx\csname url\endcsname\relax
  \def\url#1{\texttt{#1}}\fi
\expandafter\ifx\csname
urlprefix\endcsname\relax\def\urlprefix{URL }\fi
\providecommand{\bibinfo}[2]{#2}
\providecommand{\eprint}[2][]{\url{#2}}

\bibitem[{\citenamefont{Burley}(1992)}]{Burley92}
\bibinfo{author}{\bibfnamefont{R.}~\bibnamefont{Burley}},
  \bibinfo{journal}{Industrial Coating Research} \textbf{\bibinfo{volume}{2}},
  \bibinfo{pages}{95} (\bibinfo{year}{1992}).

\bibitem[{\citenamefont{Lin and Donelly}(1981)}]{Lin81}
\bibinfo{author}{\bibfnamefont{T.}~\bibnamefont{Lin}} \bibnamefont{and}
  \bibinfo{author}{\bibfnamefont{H.}~\bibnamefont{Donelly}},
  \bibinfo{journal}{A.I.Ch.E.J.} \textbf{\bibinfo{volume}{12}},
  \bibinfo{pages}{563} (\bibinfo{year}{1981}).

\bibitem[{\citenamefont{Joseph et~al.}(1991)\citenamefont{Joseph, Nelson,
  Renardy, and Renardy}}]{Joseph91}
\bibinfo{author}{\bibfnamefont{D.}~\bibnamefont{Joseph}},
  \bibinfo{author}{\bibfnamefont{J.}~\bibnamefont{Nelson}},
  \bibinfo{author}{\bibfnamefont{M.}~\bibnamefont{Renardy}}, \bibnamefont{and}
  \bibinfo{author}{\bibfnamefont{Y.}~\bibnamefont{Renardy}},
  \bibinfo{journal}{J. Fluid Mech.} \textbf{\bibinfo{volume}{223}},
  \bibinfo{pages}{383} (\bibinfo{year}{1991}).

\bibitem[{\citenamefont{Joseph}(1992)}]{Joseph92}
\bibinfo{author}{\bibfnamefont{D.}~\bibnamefont{Joseph}}, \bibinfo{journal}{J.
  Fluid Mech.} \textbf{\bibinfo{volume}{44}}, \bibinfo{pages}{127}
  (\bibinfo{year}{1992}).

\bibitem[{\citenamefont{Jeong and Moffat}(1992)}]{Jeong92}
\bibinfo{author}{\bibfnamefont{J.-T.} \bibnamefont{Jeong}} \bibnamefont{and}
  \bibinfo{author}{\bibfnamefont{H.}~\bibnamefont{Moffat}},
  \bibinfo{journal}{J. Fluid Mech.} \textbf{\bibinfo{volume}{241}},
  \bibinfo{pages}{1} (\bibinfo{year}{1992}).

\bibitem[{\citenamefont{Cohen and Nagel}(2002)}]{Cohen2002}
\bibinfo{author}{\bibfnamefont{I.}~\bibnamefont{Cohen}} \bibnamefont{and}
  \bibinfo{author}{\bibfnamefont{S.}~\bibnamefont{Nagel}},
  \bibinfo{journal}{Phys. Rev. Lett.} \textbf{\bibinfo{volume}{88}},
  \bibinfo{pages}{074501} (\bibinfo{year}{2002}).

\bibitem[{\citenamefont{Cohen et~al.}(1999)\citenamefont{Cohen, Brenner,
  Eggers, and Nagel}}]{Cohen99}
\bibinfo{author}{\bibfnamefont{I.}~\bibnamefont{Cohen}},
  \bibinfo{author}{\bibfnamefont{M.}~\bibnamefont{Brenner}},
  \bibinfo{author}{\bibfnamefont{J.}~\bibnamefont{Eggers}}, \bibnamefont{and}
  \bibinfo{author}{\bibfnamefont{S.}~\bibnamefont{Nagel}},
  \bibinfo{journal}{Phys. Rev. Lett.} \textbf{\bibinfo{volume}{86}},
  \bibinfo{pages}{1147} (\bibinfo{year}{1999}).

\bibitem[{\citenamefont{Podgorski et~al.}(2001)\citenamefont{Podgorski,
  Flesselles, and Limat}}]{Podgarski2001}
\bibinfo{author}{\bibfnamefont{T.}~\bibnamefont{Podgorski}},
  \bibinfo{author}{\bibfnamefont{J.-M.} \bibnamefont{Flesselles}},
  \bibnamefont{and} \bibinfo{author}{\bibfnamefont{L.}~\bibnamefont{Limat}},
  \bibinfo{journal}{Phys. Rev. Lett.} \textbf{\bibinfo{volume}{87}},
  \bibinfo{pages}{036102} (\bibinfo{year}{2001}).

\bibitem[{\citenamefont{Zeff et~al.}(2000)\citenamefont{Zeff, Kleber, Finberg,
  and Lathrop}}]{Zeff2000}
\bibinfo{author}{\bibfnamefont{B.}~\bibnamefont{Zeff}},
  \bibinfo{author}{\bibfnamefont{B.}~\bibnamefont{Kleber}},
  \bibinfo{author}{\bibfnamefont{J.}~\bibnamefont{Finberg}}, \bibnamefont{and}
  \bibinfo{author}{\bibfnamefont{D.}~\bibnamefont{Lathrop}},
  \bibinfo{journal}{Nature (London)} \textbf{\bibinfo{volume}{403}},
  \bibinfo{pages}{401} (\bibinfo{year}{2000}).

\bibitem[{\citenamefont{Eggers}(2001)}]{Eggers2001}
\bibinfo{author}{\bibfnamefont{J.}~\bibnamefont{Eggers}},
  \bibinfo{journal}{Phys. Rev. Lett.} \textbf{\bibinfo{volume}{86}},
  \bibinfo{pages}{4290} (\bibinfo{year}{2001}).

\bibitem[{\citenamefont{Cohen}(2002)}]{Cohen2002b}
\bibinfo{author}{\bibfnamefont{I.}~\bibnamefont{Cohen}},
  \bibinfo{journal}{Preprint Physics/0201037}  (\bibinfo{year}{2002}).

\end{thebibliography}
\end{document}